# Microwave response in a topological superconducting quantum interference device

Wei Pan[1], Daniel Soh[1], Wenlong Yu[2], Paul Davids[2], and Tina M. Nenoff[2]

[1]Sandia National Labs, Livermore, California 94551, U.S.A.
[2]Sandia National Labs, Albuquerque, New Mexico 87185, U.S.A.

**Abstract**

Photon detection at microwave frequency is of great interest due to its application in quantum computation information science and technology. Herein are results from studying microwave response in a topological superconducting quantum interference device (SQUID) realized in Dirac semimetal $Cd_3As_2$. The temperature dependence and microwave power dependence of the SQUID junction resistance are studied, from which we obtain an effective temperature at each microwave power level. It is observed the effective temperature increases with the microwave power. This observation of large microwave response may pave the way for single photon detection at the microwave frequency in topological quantum materials.

**Introduction:**

Single photon detection (SPD) has found increasingly important applications in many forefront areas of fundamental science and advanced engineering applications, ranging from studying the galaxy formation though cosmic infrared background to entanglement of superconducting qubits, single molecular spectroscopy, and remote sensing [1,2]. In recent years, the rapid developments in superconducting quantum computation, high fidelity quantum measurement, quantum key distribution, and quantum network call for SPD in the microwave frequency range [3]. The current SPD scheme has good sensitivity for photons in the high frequencies range (e.g., visible light). However, their sensitivity decreases drastically for low-frequency, low energy, microwave photons. As a result, the detection of single photons at this low frequency is highly prone to error from classical noise.

Graphene single photon detectors (i.e., graphene superconducting Josephson junctions) have emerged as one new platform to meet the needs of detecting single microwave photons [4,5]. It is capable of performing SPD over a wide frequency range, particularly at the infrared and microwave frequencies due to its linear energy dispersion relationship. Like graphene, the helical surface states in $Cd_3As_2$, a Dirac semimetal [6-8], also possess Dirac linear dispersion relationship. As a result, $Cd_3As_2$ is also sensitive to low-frequency microwave photons. Compared to graphene, $Cd_3As_2$ may be even more promising for microwave photon detection [9] based on the following reasons. First, a higher electron mobility has been reported. Indeed, a mobility as high as $10^7$ cm$^2$/Vs has recently been reported in Dirac semimetal $Cd_3As_2$ single crystals [10]. Second, they can be readily grown by many conventional growth techniques, such as vapor transport [11], MBE [12], PLD [13] techniques; this enables their facile integration into any optical device structures, such as microwave cavities. Third, the unique electronic and optical properties in $Cd_3As_2$ may allow for polarization-resolved photon detection [14]. Fourth, superconductivity in $Cd_3As_2$ thin films [15] and the supercurrent states in $Cd_3As_2$-based Josephson junctions via the superconducting proximity effect [16-18] have been demonstrated, receptively. This may make the adoption of the well-developed single photon detection schemes, such as superconducting nanowires and transition edge sensors [2], possible in the $Cd_3As_2$ material system. Final, the helical surface states in topological semimetals, when combined with conventional superconductors, can host Majorana zero modes, which can be used to construct topological qubits. New single photon detection scheme utilizing Majorana zero modes have also been proposed recently [19]. Together, the microwave single photon detection capability and qubit operation is predicted to lead to high-fidelity quantum computation [20].

In this paper, microwave response in this proximity induced superconducting state is presented in a superconducting quantum interference device (SQUID) structure fabricated on $Cd_3As_2$, as shown in the inset of Figure 1(a). In our SQUID device, a large photo response is observed at various microwave frequencies ranging from 0.5 to 10 GHz.

**Device and Methods:**

The mechanical exfoliation method is used to obtain flat and shiny $Cd_3As_2$ thin flakes from the initial bulk ingot materials. Information about the $Cd_3As_2$ polycrystalline ingots can be found in Ref. [16]. The thickness of the resulting exfoliated $Cd_3As_2$ flakes is approximately 200nm. To fabricate the alminimum-Cd$_3$As$_2$-almuninum SQUID, a two-step process is employed. First, a Cd$_3$As$_2$ flake is deposited on a Si/SiO$_2$ substrate (with SiO$_2$ thickness of 1 μm). Then, e-beam lithography is used to define the aluminum (Al) electrodes. The thickness of resultant Al electrodes is 300 nm. A low-frequency (~ 11 Hz) phase-sensitive lock-in amplifier technique, with an excitation current of 10 nA, is used to measure the sample resistance. To measure the differential resistance, a large direct current (up to ± 2μA) is added to the 10 nA *a.c.* current. The entire device is immersed in cryogenic liquid; all measurements are carried out at the cryogenic temperature of ~ 0.25K.

**Results and Discussion:**

Figure 1a shows the temperature dependence of the SQUID resistance $R_{xx}$. At high temperatures, the $R_{xx}$ is nearly constant. The drop at T ~ 1.2K is due to the onset of the superconductivity in the aluminum electrodes. $R_{xx}$ continues to decrease slowly from 1.2K to ~ 0.55K. After 0.55K, $R_{xx}$ drops precipitately and reaches a zero-resistance state at T ~ 0.35K. We thus take 0.55K as the superconducting transition temperature ($T_c$). Direct current-voltage (I-V) measurements in this junction is shown in Figure 1b. For large *d.c.* currents $I_{DC}$, the I-V curve follows a linear dependence. From the slope of this straight line, a normal state resistance of $R_n \approx 75$ Ω can be deduced. Extrapolating the line to zero $V_{dc}$, we obtain an excess current of ~ 0.08 μA. Assuming the two Josephson junctions in the SQUID are identical and taking into account the superconducting gap of $\Delta = 1.75 k_B T_c$, we can estimate the barrier strength Z ~ 1 in our SQUID, based on the calculations in the paper by Flensberg et al [21,22]. Correspondingly, the junction transparency T = 1/(1+Z$^2$) is estimated to be ~ 0.5. In the small $I_{DC}$ regime of |$I_{DC}$| < 1 μA, the voltage across the junction $V_{dc}$ is zero, demonstrating the robust supercurrent states. The critical current is approximately 1μA.

Figure 1c shows the magnetic field dependence of the I-V data in this SQUID. Periodic oscillations of the critical current are clearly seen, as expected for a conventional SQUID. The period is estimated to be ~ 1.8 mT. This corresponds to an effective SQUID area of ~ 1.1 μm$^2$, as illustrated by the dashed square in the inset of Fig. 1a. We note that in a small SQUID the effective area is often larger than the middle open area, due to the flux compression effect by the surrounding electrodes [23]. Another feature in Fig. 1c is the envelop of the oscillatory pattern being modulated by the Fraunhofer diffraction pattern of the single Josephson junction in the SQUID.

To examine the microwave response in our SQUID device, a setup shown schematically in Figure 1d is utilized. An Agilent 83592B sweep generator is used to generate microwave photons, which are conducted through a semirigid coax cable. The end of coax cable is located about 5 mm above the sample surface. The microwave power is tuned at room temperature; the exact microwave power at the end of coax cable is not known.

The differential resistance dV/dI as a function of *d.c.* current bias ($I_{DC}$) in this SQUID is measured and reported in Figures 2a, 2b, and 2c; it has been collected at the three selected microwave frequencies of 0.5, 7, and 10 GHz, respectively. Additionally, the microwave power was varied at each frequency. The SQUID shows a large response at both the zero and also the finite *d.c.* bias. In this paper, we will focus on the microwave response at the zero-bias current. Other features, such as the Shapiro steps [24] at non-zero *d.c.* bias, merits more detailed studies and will be discussed elsewhere.

A general trend is seen when the resistance as a function of microwave power for each microwave frequency is plotted, see Figures 2d-f. At low microwave powers, the resistance is approximately zero. With further increase in microwave power, resistance becomes non-zero and starts increasing. We note here that the onset microwave power (in dBm) for non-zero dV/dI is different for different microwave frequencies, e.g., -33 dBm for f = 0.5 GHz and -16 dBm for f = 7 GHz. This is mainly due to different power attenuations at different frequencies through the semi-rigid coax cable we used in the measurements.

Overall, the microwave power dependence of the zero-bias resistance shows a similar trend to its temperature dependence, seen in Fig. 1a. The zero-bias resistance is zero at low microwave powers (versus zero resistance at low temperatures). After an onset microwave power, the zero-bias resistance increases with increasing microwave power. This, again, is like the temperature dependence, in which the resistance increases with increasing temperature after a critical temperature. The similarity suggests that the increase of zero bias resistance probably is of a bolometric origin, *i.e.* due to the increasing temperature upon microwave photon absorption. Based on this assumption, it is possible to deduce the effective electron temperature at various microwave powers by relating the resistance value with the temperature dependence from Fig. 1a.

It is important to note that this procedure works well for mid-range microwave powers, where the resistance is finite and has not reached the normal state resistance. For lower microwave powers, the resistance is zero thereby rendering it difficult to determine the effective electron temperatures. For higher microwave powers, or higher electron temperatures, the resistance change is gradual, and the error bar is relatively large. In Figure 3, a plot shows the effective electron temperature as a function of mid-range microwave power (in a logarithmic scale). It is clearly shown that the effective temperature increases roughly linearly with increasing microwave power (in the logarithmic scale).

The above result, i.e., effective temperature increasing with increasing photon energy, is promising for microwave photon detection. Indeed, with further optimization of microwave coupling structure, for example through utilization of a meander line [25], quarter wave resonator [4], or log periodic antennas [26], measurements can be done at much lower microwave power level, which may provide more support for single photon detection with number resolving [9, 27] capability. Compared to other superconducting photon detectors, such as transition edge sensors (TESs), the photon detection in this device is done by the zero-bias resistance, thus avoiding a large source-drain current needed, for example, in a TES structure [28]. Consequently, issues caused by the large source-drain current, such as the flicker noise, are greatly reduced.

**Conclusion:**

In summary, a large microwave response has been observed in a superconducting quantum interference device fabricated on Dirac semimetal $Cd_3As_2$ thin flakes, in which the temperature dependence and microwave power dependence of the junction resistance are studied. The effective temperature of the junction device under microwave radiation increases with increasing microwave power (in the logarithmic scale). This result may pave the way of single photon detection at the microwave frequency in topological quantum materials.

**Acknowledgements**

The work was supported by a Laboratory Directed Research and Development project at Sandia National Laboratories. Device fabrication was performed at the Center for Integrated Nanotechnologies, an Office of Science User Facility operated for the U.S. Department of Energy (DOE) Office of Science. Sandia National Laboratories is a multimission laboratory managed and operated by National Technology & Engineering Solutions of Sandia, LLC, a wholly owned subsidiary of Honeywell International Inc., for the U.S. Department of Energy's National Nuclear Security Administration under contract DE-NA0003525. This paper describes objective technical results and analysis. Any subjective views or opinions that might be expressed in the paper do not necessarily represent the views of the U.S. Department of Energy or the United States Government.

**Figures and figure captions:**

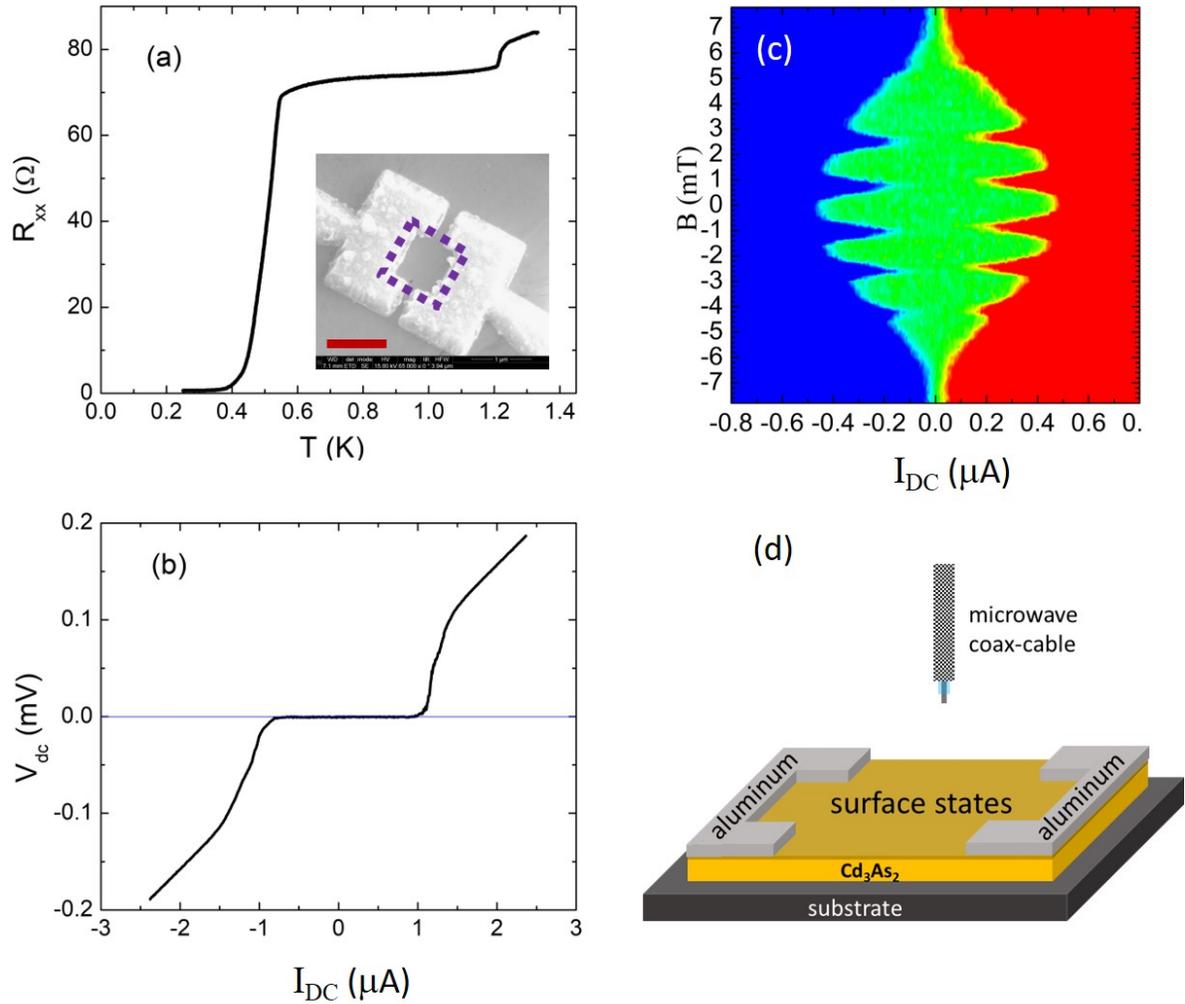

Figure 1: (a) The temperature dependence of the junction resistance in a superconducting quantum interference device (SQUID). The insert shows the SQUID device fabricated on a $Cd_3As_2$ thin flake. Th scale bar is 1 µm. (b) The current-voltage (I-V) curve measured in the SQUID. The critical current is ~ 1 µA. (c) The two-dimensional color plot of I-V traces as a function of magnetic fields at 0.44K. The red color represents a positive $V_{dc}$, blue for negative $V_{dc}$. The green area represents the supercurrent regime. (d) The schematic setup (dimension not to scale) used to examine microwave response.

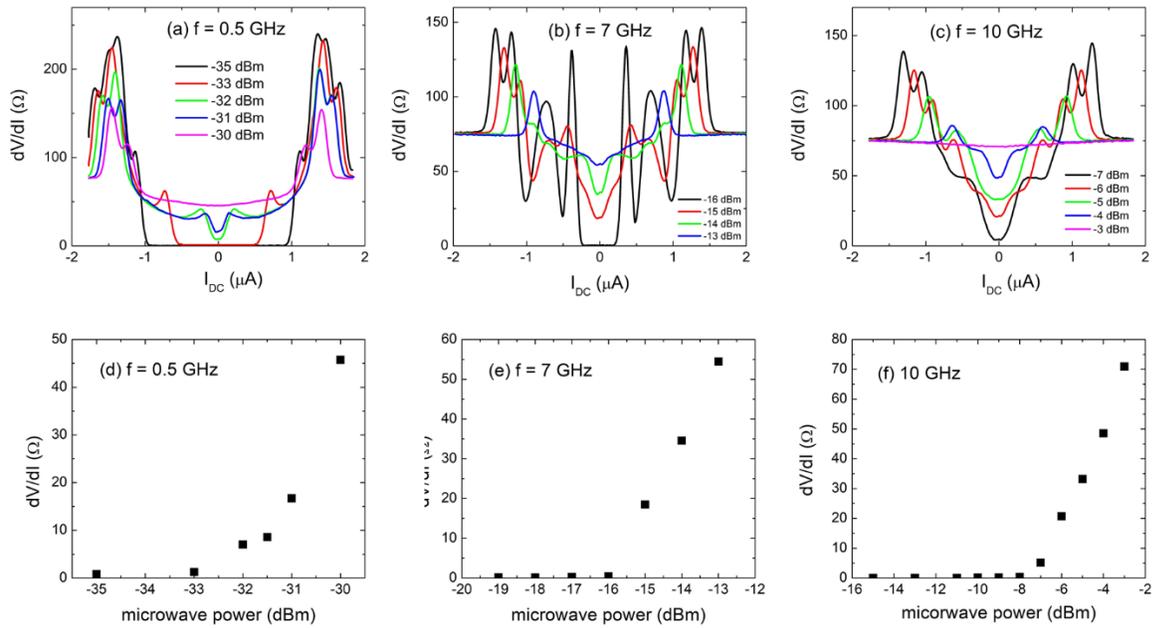

Figure 2: (a)-(c) The differential resistance measured at three selected microwave frequencies, 0.5, 7, and 10 GHz. At each frequency, the microwave power is also varied. (d)-(f) The zero-bias resistance as function of microwave power at 0.5, 7, 10 GHz, respectively.

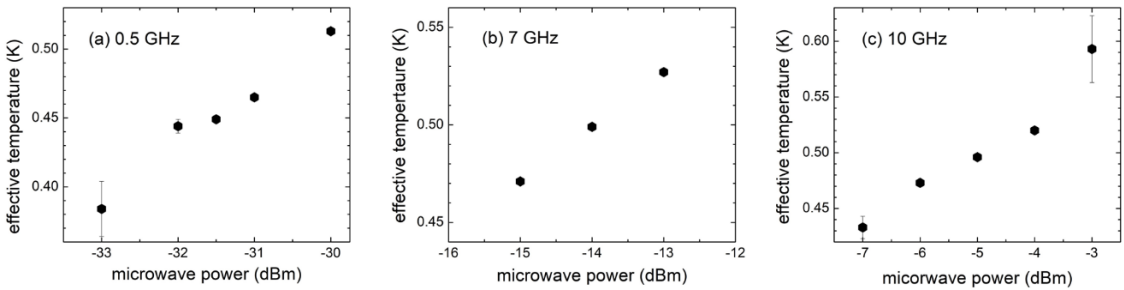

Figure 3: The effective junction temperature as a function of microwave power at 0.5 (a), 7 (b), and 10 (c) GHz.